# The Effectiveness of Classification on Information Retrieval System (Case Study)

**Maher Abdullah and Mohammed GH. I. AL ZAMIL**

Department of Computer Information Systems

Yarmouk University, Jordan

[Mohammedz@yu.edu.jo](mailto:Mohammedz@yu.edu.jo)

## Abstract

Large amount of unstructured designed information is difficult to deal with. Obtaining specific information is a hard mission and takes a lot of time. Information Retrieval System (IR) is a way to solve this kind of problem. IR is a good mechanism but does not give the perfect solution. Other techniques have been added to IR to develop the result. One of the techniques is text classification. Text classification task is to assign a document to one or more category. It could be done manually or algorithmically. Text classification enhances the output of this process by reducing the results. This study proved that text classification has a positive influence on Information Retrieval Systems.

# 1. Introduction

Obtaining the necessary information at the right time from a large amount of data sources is the goal for information seekers. In fact, extracting the required information from documents that include important data is critical task. Large institutions have their own hierarchies, procedures and regulations, which are subject to expand and change frequently. On the other hand, different managements levels need to get timely and accurate information as quick as possible [18].

Information Retrieval (IR) systems are candidate solution for handling such task. However, accuracy of retrieved data is still an emergent issue since many applications are financially critical or vitally severe. The main reason behind this low performance is that information retrieval usually deals with natural language text which is unstructured and could be semantically ambiguous; unlike structured designed data retrieval systems (e.g. Rational Database Systems), which deal with data that have a well-designed semantic and structure [8, 19, 20].

The users of IR systems must interpret and translate their information needs into keywords or queries by the language description provided by the system they use; usually natural language. With IR systems, this implies specifying a set of words which convey the semantic of information need. But characterization of the user information need is not a simple problem [21].

The organization and representation of information items should provide users with easy access, secure, and accurate information in which they are interested to retrieve. IR systems such as Yarmouk system, which deal with information that is redundant and vast, make it hard to the users to find the necessary information. This type of systems can be described as follows [26, 27]:

1.  *Information overload:* the vast amount of information is a big challenge that faces this kind of systems but sometimes the challenge is the redundancy that documents include. Finding the pertinent information is considered a critical issue for the user.

2.  *Time consuming*: users need to find the necessary information as quickly as possible so that the users do not waste time in only one or few tasks.

To deal with such unstructured information in which the amount of data is huge and growing up dramatically, you need to build an information retrieval system with good searching capabilities which will be did in this project.

The objective of this project is to estimate the influence of text classification on information retrieval system. Traditional information system will be established and the result the system show will be estimated. Adding text classification to the indexing in the system will affect the result that show. This implementation will enhance the search for information result.

In addition, the implementation could be used in Yarmouk University System to help manager and employees to get better information services by enabling them to obtain the information they need on time.

## 2. Literature Review

Information Retrieval (IR) is the discipline of acquisition information relevant to needed information from a collection of information resources [3]. IR deals with organize, storage, represent and access unstructured and semi-structured information records such as documents, online catalogues, webpages and multimedia objects. Recently, there are noticeable needs for effective techniques that automate the process of information retrieval since most of data sources provide unstructured pool of datasets. [11].

Since the volume of information is dynamically growing, it is necessary to build a specialized data structure for fast search, which called index. Indexes are core for every modern information retrieval system. Indexes enable a fast access to the data and allow speed up of query processing. Then, the retrieval processing comes to produce the hit list, which composed of retrieving documents that satisfy a user query [13].

### 2.1. Information Retrieval

IR systems consist mainly of building up effective indexes that organize item-data into an organized fashion, processing user queries and developing ranking algorithms to enhance the results by displaying most interested ones. The first phase of establishing an IR system is to gather the document collection and store it in repository, which formulate the corpus of the system. In the second phase, the documents need to be organized and indexed for fast retrieval and ranking. The most popular indexing structure is inverted index, which consists of all the distinct words in the corpus and for each word in it a list of linked that pointed to documents that contain it [1, 5].

An inverted file (inverted index) is a word-oriented technique for indexing a collection of text to speed up the searching task. The structure of inverted files is composed of two elements: the vocabulary and occurrences. The vocabulary is the set of all different words (terms) in the text. For each term a list of all the text occurrences. Figure 2.2.1 illustrate inverted index [20]. Other indexing could be used such as suffix trees and suffix arrays.

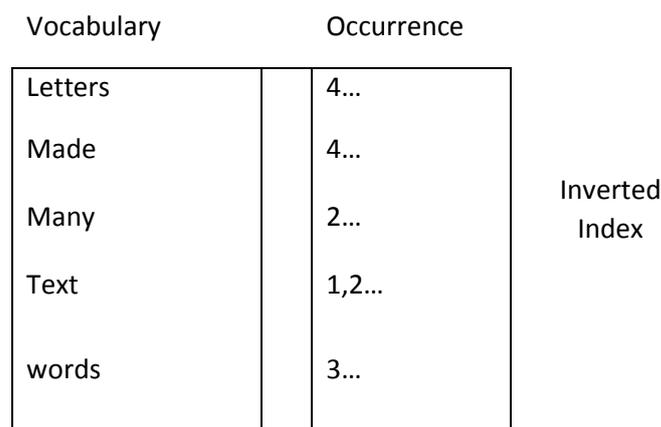

Figure 1. Inverted Index

(source: Baeza-Yates & Ribeiro-Neto, 2004)

IR typically seeks to find documents in each collection that are about a specific topic or that satisfies a given information need. The user expresses needed topic or information by generating a query. Documents that satisfy the expressed query - in the judgment of the user - described as "relevant". Documents that are not related to the given query descried as "non-relevant". One major function of an IR engine is to classify documents into their designated clusters for facilitating the way the results displayed to the users. Naturally, classification of the documents plays an important role in the enhancing of the result [9, 10]. In other words, the better the classification features, the higher the proportion of documents returned to the user that will be judged as relevant [4, 6].

Researchers mentioned that indexing is very important step to enhance the effectiveness and efficiency of the search [11]. Index is a data structure that built on a text to speed up the search. As shown in figure 2, indexing could go through many stages

1. Stop-word removal: remove the words that are redundant in every document.
2. Stemming: return each word to its grammatical root to reduce distinct words.
3. Remove space and symbols (e.g. -, _, *, ?)
4. Noun grouping: put nouns in groups to classify them.

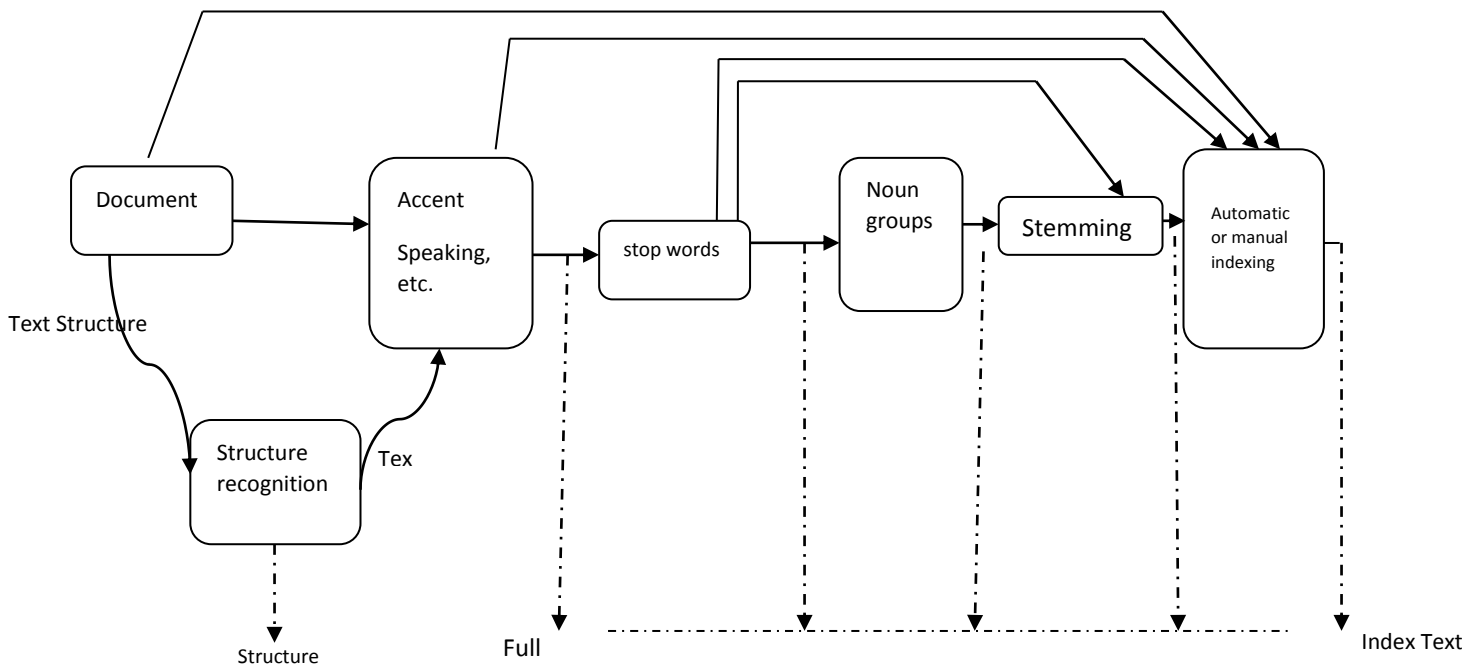

Figure. 2 indexing stages

Or it could be full-text indexing. Now system is ready to be searched in it. In the selection stage, to rank your result there several methods that could be used:

1. Term documents frequency: a simple and effective term selection procedure is to use only the terms whose document frequency exceeds a pre-established frequency threshold to represent the documents.

2. TF-IDF weights: similar but slightly more sophisticated procedures to adopt a term selection procedure that retains the terms of higher TF-IDF weights in each document.

To measures the success of IR, two concepts of relevance that are widely used: *precision*, which defined as "the ratio of relevant items retrieved to all items retrieved, or the probability given that an item is retrieved that it will be relevant" and recall, which is defined as "the ratio of relevant items retrieved to all relevant items in the collection, or the probability given that an item is relevant that it will be retrieved" [5].

As a final step, IR engines reordered the hit set of documents according to a given ranking features. For instance, if a document D1 received higher ranking than D2, this implies that D1is likely to be more relevant to a given user query than D2 [3, 8].

Measuring how relevant. IR is fundamentally concerned with r each document to a given query is an important issue too. Retrieving information that is relevant to a user's need or query requires matching each document with a given query. Some of IR systems to deal with the subjectivity of relevance, they need to generate user profiles, i.e., user preferences. The objective of such profile is to give extra information according to predefined criteria [14, 16].

Other approach to enhance the result of retrieved documents is to classify the collection. Putting each group of the collection that talking about the same general label in a single name for the whole group [28, 29].

## 2.2 Text Classification

First tactic for categorizing documents is to assign a label to each document, but this solve the problem only when the users know the labels of the documents they looking for. This tactic does not solve more generic problem of finding documents on specific topic or subject. For that case, better solution is to group documents by common generic topics and label each group with a meaningful name. Each labelled group is called category or class [2, 7].

Document classification is the process of categorizing documents under a given cluster or category using fully supervised learning process. Classification could be performed manually by domain experts or automatically using well-known and widely used classification algorithms such as decision tree and Naïve Bayes [15].

Documents are classified according to other attributes (e.g. author, document type, publishing year etc.) or according to their subjects. However, there are two main kind of subject classification of documents: The content based approach and the request based approach. In Content based classification, the weight that is given to subjects in a document decides the class to which the document is assigned. For example, it is a rule in some library classification that at least 15% of the content of a book should be about the class to which the book is assigned [16]. In automatic classification, the number of times given words appears in a document determine the class. In Request oriented classification, the anticipated request from users is impacting how documents are being classified. The classifier asks himself: "Under which description should this entity be found?" and "think of all the possible queries and decide for which ones the entity at hand is relevant" [12].

Automatic document classification tasks can be divided into three types [15, 22]:

1. Unsupervised document classification (document clustering): the classification must be done totally without reference to external information.
2. Semi-supervised document classification: parts of the documents are labelled by the external method.
3. Supervised document classification where some external method (such as human feedback) provides information on the correct classification for documents

## 2.3 Classifier

There are several methods that can be used in text classification [12]. K-Nearest Neighbours (KNN) algorithm relies on measuring the distance between a cluster centroid and every coming object. The lower the distance implies more relevancies. Naïve Bayes classification algorithm is designed on the top of a probabilistic framework, in which the probability that a given object is likely to belong to specific category is computed. The higher the probability is the more chance to assign the object to a given cluster. Decision tree structure simulates a divide and conquers approach, where features are used to build a decision tree and leaf are representing class labels or categories [24].

## 2.4 Clustering

Document clustering is the process of reorganizing documents so that documents that are related semantically are grouped into a similar cluster. The operation of clustering documents is usually of two types [23]:

1. Global clustering strategy: The documents are grouped accordingly to their occurrence in the whole collection.
2. Local clustering strategy: the grouping of documents is affected by the context defined by the current query. And its local sets of retrieved documents.

Clustering methods are usually used in IR to transform the original query in attempt to better represent the user information need. From this perspective, clustering is an operation which is more related to the transformation of the user query than the transformation of the text of the documents [25].

## 3. Methodology

IR systems are dealing with unstructured designed documents like our case [30, 31]. They help users to get information they need with minimum time lost. In this study, the author tries to accelerate the process of getting the needed information by adding classification to the indexing. As shown in figure 3.1, in this study the author tries to make a comparison between traditional IR system and IR system with classification.

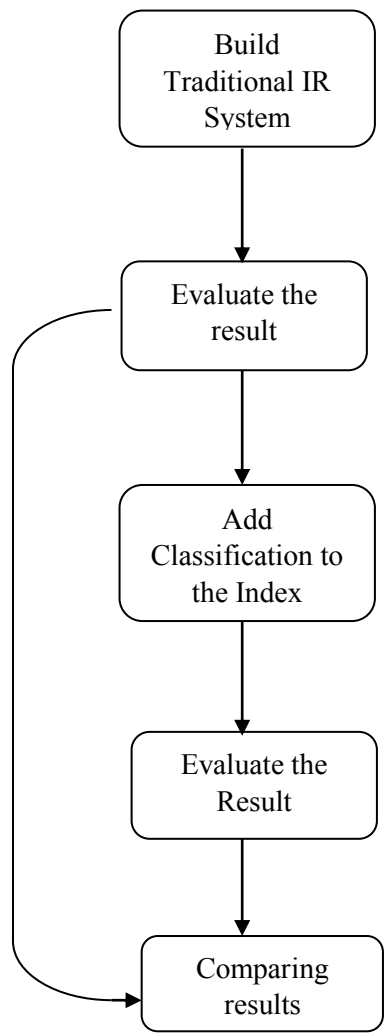

Figure 3.1: Methodology

The used corpus in this project was the laws and regulation of Yarmouk University which the author has got it from Department of Legal Affairs. The document is classified to ten categories:

1. Higher Education and Jordanian universities: contains six documents.

2. Yarmouk University: contains nineteen documents.

3. Teaching Staff at the University: contains fourteen documents.

4. Personnel at the University: contains seventeen documents.

5. Students: contains twelve documents.

6. Public Financial Affairs and Supplies: contains eighteen documents.

7. Financial Affairs of Students: contains three documents.

8. Degrees and Certifications: contain nine documents.

9. Scientific affairs and research: contains eleven documents.

10. Scientific Centres, Institutes and Schools: contains fifteen documents.

In fact, this same classification will be used later in the implementation at second phase when creating indexing with classification which will be explained later in this chapter.

## 3.1. Traditional IR

At first phase in the methodology, traditional IR system has been built and the four steps in the figure 3.1 have been processed. Step one is building the index which explained in figure 3.1.1. The figure shows that the program accesses each file in the corpus and read it word by word to the end. The program takes each word and stores it in the database with its file-name, path and frequency. Stemming have been escaped from indexing process because researchers mentioned that it has negative effect in Arabic text index [21].

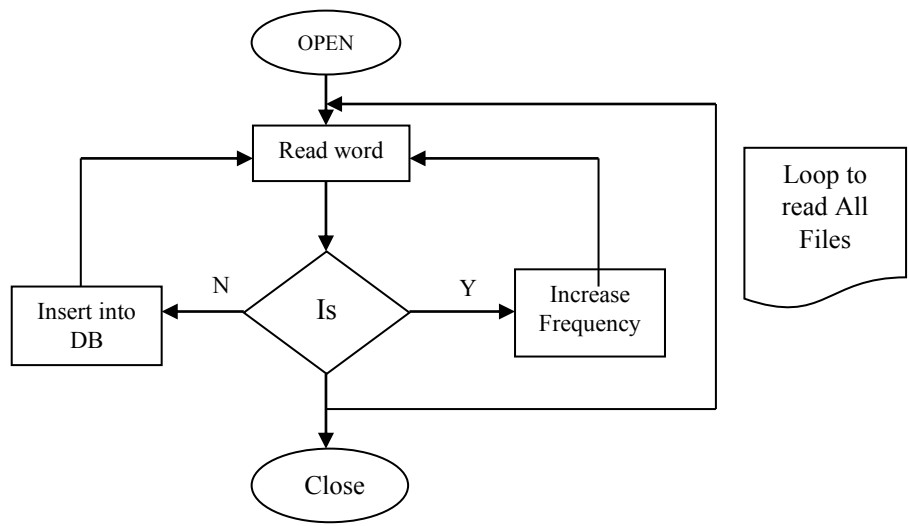

Figure 3.1.1: building the index

The user translates its need to a query and search for it. The program matches terms in query with the indexing and return the result to the user ranked according to the summation of term frequency [32, 33, 34]. The ranking considers finding all the term of the query into the document then the summation of the term frequency. From the result, the user selects the relevant ones and from applied the recall and precision equation the result will be evaluated.

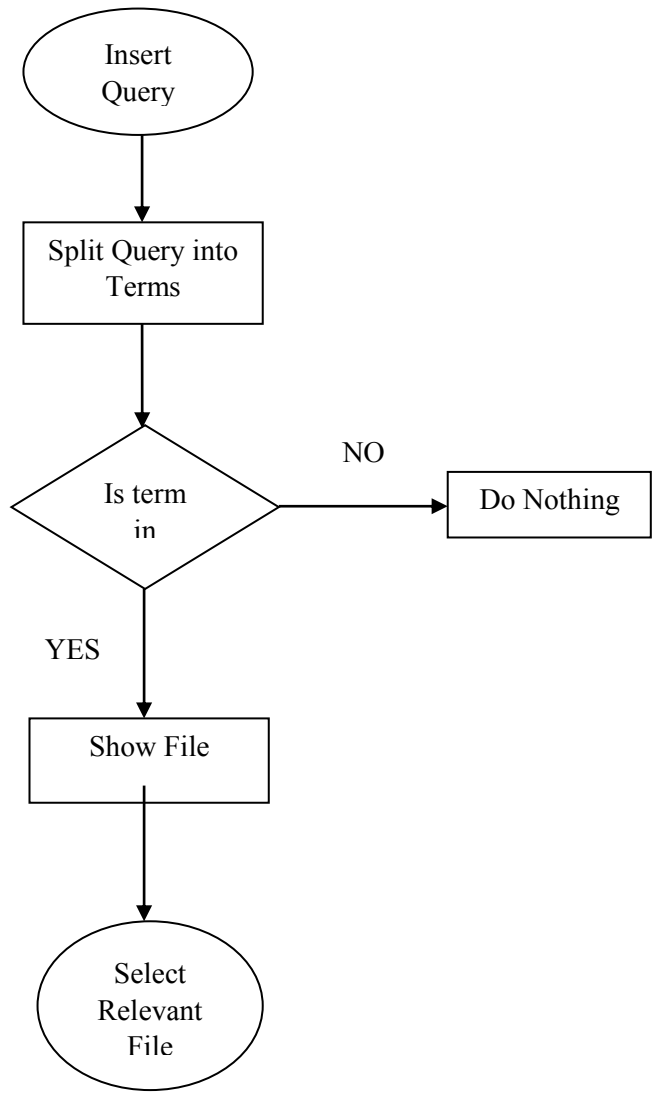

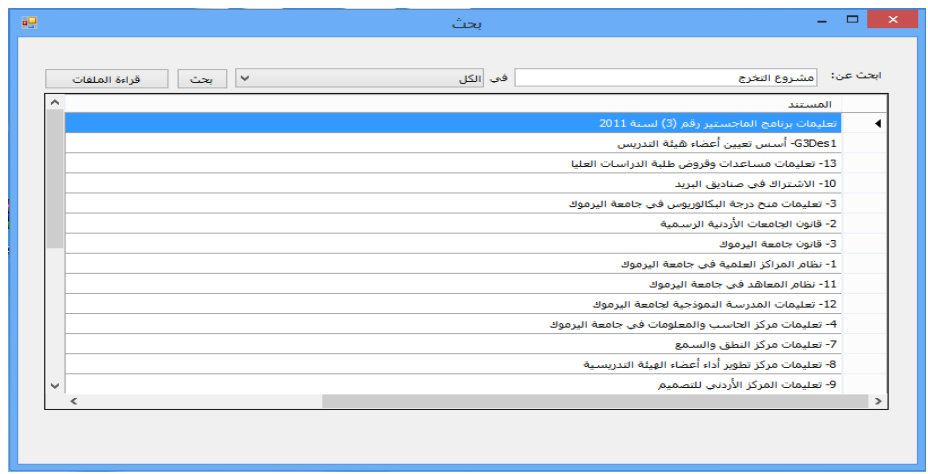

Figure 3.1.3: Sample of Result before

## 3.2 IR with Classification Phase

In this phase of the methodology, classification is added to the index. The classification of documents that mentioned earlier in this chapter is classified by Department of Legal Affairs and they are available on Website of Yarmouk University (www.yu.edu.jo). Classification that has added to the index changes the schema of the database to be as shown in figure 3.2.1.

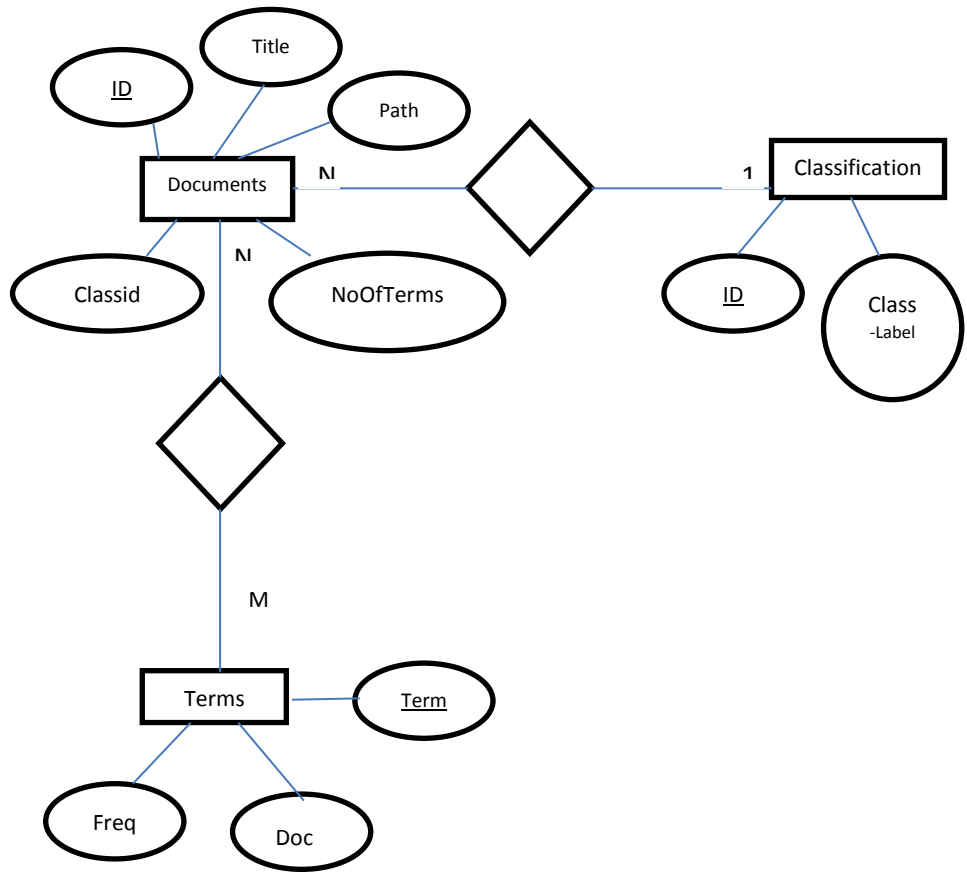

Figure 3.2.1: Schema of Database after Classification

The rest of steps are like the step in first phase which translate the needed information into query and make matching to show the result to the user and start to select the relevant documents.

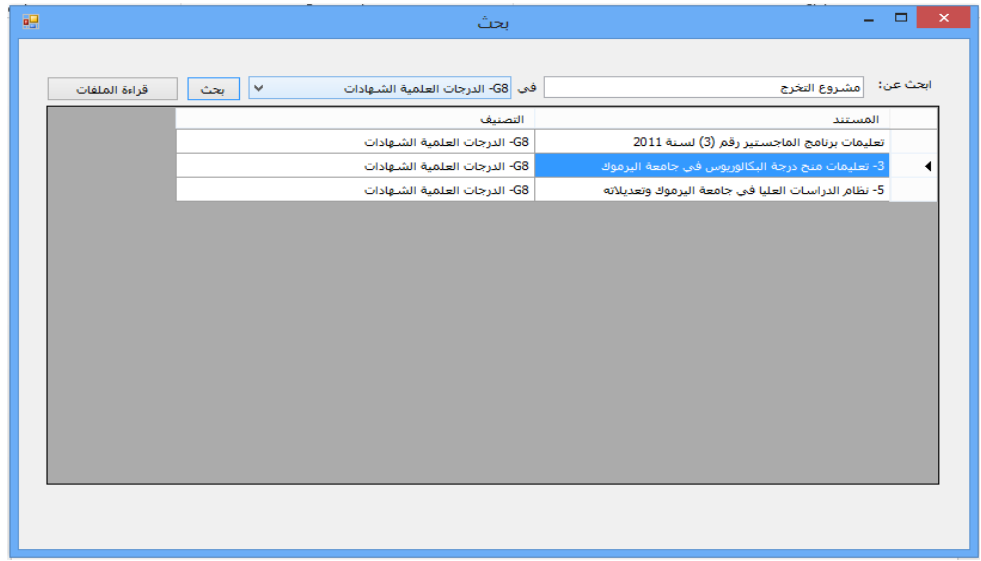

Figure 3.2.2: Sample of Result before Classification

## 4. Experiments and Evaluation

The implementation has been made by visual studio.net using C#.NET language. First to build the index, the user need to show the implantation the place of the collection of files to be read from as figure 4.1 shows.

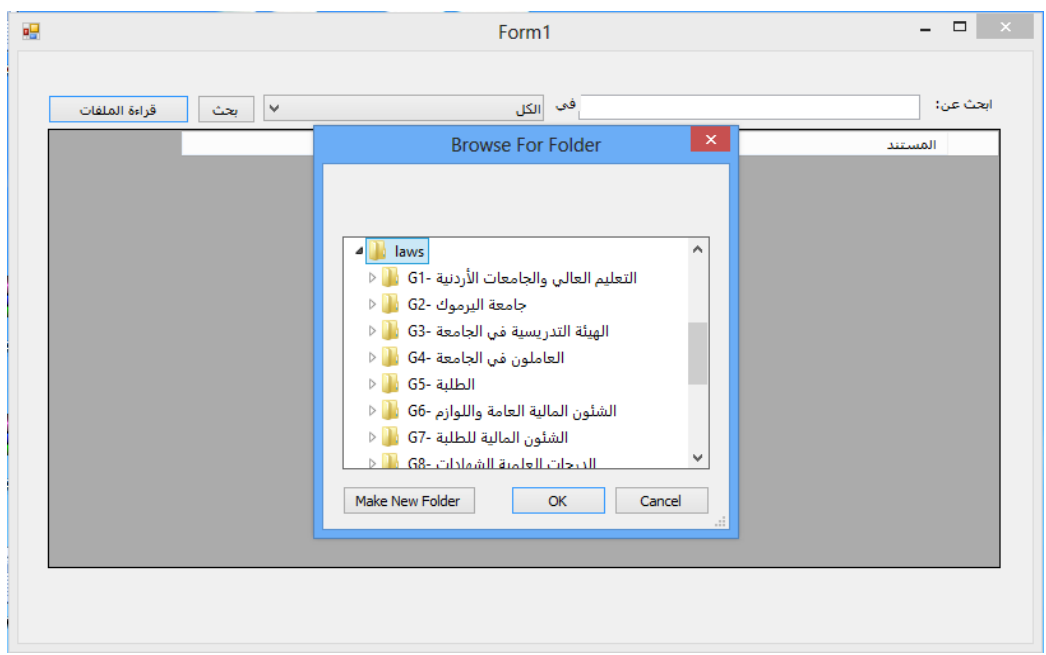

Figure 4.1: locate the path of corpus

Then, the implementation accesses each file in each sub-directory and read word by word to store it in the index. If the word is already existing, the frequency if the word increase by one. The implementation also need to store the path of each document to enable the user to open that documents and read from it, see appendix A. now the implementation is ready for the

user to search. In the search process, the implementation goes throw steps that shows in figure 3.1.2

## 4.1 Results

Researchers usually use one or more measurements to measure the impact of classification on the search process. Table 4.1.1 shows samples of the result the author gets from implementing the application. The measurements have been used here is precision and recall which is the most popular measures in IR field. The equations that we used to calculate precision and recall are:

$$\text{recall} = \frac{|\{relevant\ document\} \cap \{retrieved\ documents\}|}{|\{relevant\ document\}|}$$

And

$$\text{Precision} = \frac{|\{relevant\ document\} \cap \{retrieved\ documents\}|}{|\{retrieved\ document\}|}$$

**Table 4.1.1 results**

| QUERY | PRE $_{before}$ | RECALL $_{before}$ | PRE $_{after}$ | RECALL $_{after}$ |
|---|---|---|---|---|
| بعثات | 1/3 | 1 | 1 | 1 |
| التأمين الصحي | 2/12 | 1 | 2/4 | 1 |
| الرحلات الجامعية | 1/23 | 1 | 1/4 | 1 |
| معادلة الشهادات | 1/4 | 1 | 1 | 1 |
| تثبيت أعضاء الهيئة التدريسية | 1/62 | 1 | 1/23 | 1 |
| معاملة ترقية | 1/17 | 1 | 1/6 | 1 |
| ميزانية الجامعة | 1/109 | 1 | 1/10 | 1 |
| التعيينات | 1 | 1 | 1 | 1 |
| تشكيل المجالس في الجامعة | 1/125 | 1 | 1/14 | 1 |

## 4.2 Discussion

From table 4.1.1, you can see the difference in result before and after using the classification and how classification affects the result. As you can see in query "بعثات", before the classification, the application retrieved three documents but only one of them was relevant. After the classification the application retrieved one document which was the relevant one. This is an example of query with one word. Next three examples in the same table each query is consisted of two words. Frist of them, "التأمين الصحي" which before classification the application retrieved twelve documents only two of them are relevant and after classification the application retrieved only four and two of them are relevant. The second query of the group is "الرحلات الجامعية" which as table 1 shows, the application retrieved twenty-three documents and one of them is relevant, but after classification, the application retrieved four documents and one of them is relevant. The last query that have two words is "معادلة الشهادات" which before classification the application retrieved four documents, only one of them is relevant. And after classification, the application retrieved only one document which was the relevant one.

The last query shown in table 1 is a query with four words "تثبيت أعضاء الهيئة التدريسية". As shown in table 1, the application retrieved sixty-two documents only one of them is relevant, and after classification, the application retrieved twenty-three documents only one of them is relevant.

As it is clearly from table 1, recall is always 1 which means the users considered that the relevant documents are always shows off between the retrieved ones. This is also means that users are not specialist in the regulation and laws of the university (the corpus), but it also means there is no confect between these laws and regulations. And that make the recall not good measure in this situation. And the result in table 1 shows that there is a big gap between the values of precision measure. The smallest value is 1/125 (0.008) and the biggest value is 1. This mean choosing the keyword of the query is make a difference.

From column 2 (precision before classification) and column 4 (precision after classification), the gab is big which means classification have a big positive influence on IR systems

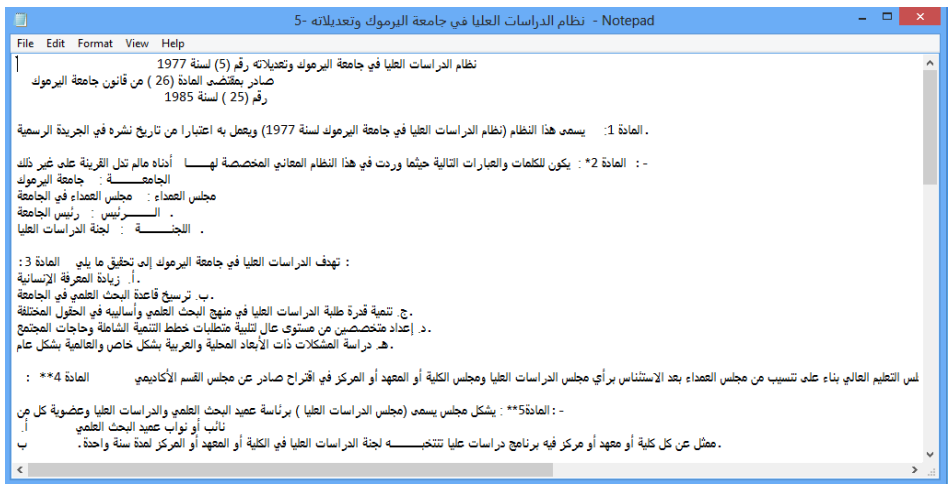

Figure 4.2.1 shows a sample of the documents that the program is dealing with.

## 5. Conclusion

IR systems solved the problem of finding specific information form a lot amount of unstructured information. However, IR is not always the perfect solution for this kind of problem to researchers found some techniques helped enhancing the result that IR systems come with. Text classification is one of those techniques.

From table 1 you can see clearly that classification has a positive impact on the search. Classification make the number of retrieved documents reduce in some cases to more than half. This reflected positively on the value of precision which means the search has been enhanced. This enhancing reduces the time that the users need to spend in the retrieved documents to get what he/she wants.

As mentioned earlier researchers claimed that stemming have a negative influence in Arabic text indexing, the future work will measure the stemming with classification in indexing.